\documentstyle[namedreferences,psfig]{kapproc}  

\def\references{\bibliographystyle{swanbib}\bibliography{kris}}
 
\def\barecite#1{\citeauthor{#1}, \citeyear{#1}} 

\renewcommand{\cite}[1]{\citeauthor{#1} (\citeyear{#1})}
\renewcommand{\[}{\begin{equation}}
\renewcommand{\]}{\end{equation}}

\newcommand{\vc}[1]{\mbox{\bf #1}}
\newcommand{\mx}[1]{\hat{#1}}

\newcommand{\tfrac}[2]{{\textstyle\frac{#1}{#2}}}

\newcommand{\GB}{\langle|\vc B|\rangle}
\newcommand{\NB}{\langle\vc B\rangle}
\newcommand{\GBx}[1]{\langle| B_{#1}|\rangle}
\newcommand{\NBx}[1]{\langle B_{#1}\rangle}
\def\arcsec{\hbox{$^{\prime\prime}$}}
\def\ptl{\partial}
\def\epsi{\epsilon}
\def\lang{\langle}
\def\rang{\rangle}
\def\la{\mathrel{\mathchoice {\vcenter{\offinterlineskip\halign{\hfil
 $\displaystyle##$\hfil\cr<\cr\sim\cr}}}
 {\vcenter{\offinterlineskip\halign{\hfil$\textstyle##$\hfil\cr
 <\cr\sim\cr}}}
 {\vcenter{\offinterlineskip\halign{\hfil$\scriptstyle##$\hfil\cr
 <\cr\sim\cr}}}
 {\vcenter{\offinterlineskip\halign{\hfil$\scriptscriptstyle##$\hfil\cr
 <\cr\sim\cr}}}}}

\runningtitle{PASSIVE MAGNETIC FIELD TRANSPORT} 
\begin{opening}
  \title{Theory of passive magnetic field transport} 
  \author{Krist\'of Petrovay}
  \institute{E\"otv\"os University, Department of Astronomy \\
             Budapest, Ludovika t\'er 2, H-1083 Hungary}
\end{opening}

\begin{document}
\footnotetext{In: \it Solar
      Surface Magnetism, \rm (R. J. Rutten \& C. J. Schrijver, eds.), 
      NATO ASI Series C433, Kluwer 1994, p.\ 415-440. 
      }
\begin{abstract}
In recent years, our knowledge of photospheric magnetic fields went through 
a thorough transformation---nearly unnoticed by dynamo theorists. It is now
practically certain that the overwhelming majority of the unsigned magnetic 
flux crossing the solar surface is in turbulent form (intranetwork and hidden 
fields). Furthermore, there are now observational indications (supported by 
theoretical arguments discussed in this paper) that the net polarity imbalance 
of the turbulent field may give a significant or even dominant contribution to 
the weak large-scale background magnetic fields outside unipolar network areas. 
This turbulent magnetic field consists of flux tubes with magnetic
fluxes below $10^{10}$\,Wb ($10^{18}$\,Mx). The motion of these thin tubes is
dominated by the drag of the surrounding flows, so the transport of this 
component of the solar magnetic field must fully be determined by the 
kinematics of the turbulence (i.e.\ it is ``passive''), and it can be 
described by a one-fluid model like mean-field theory (MFT). The recent 
advance in the direct and indirect observation of
turbulent fields is therefore of great importance for MFT as these are the 
first-ever observations on the Sun of a field MFT may be applied to. However,
in order to utilize the observations of turbulent fields and their large-scale
patterns as a possible diagnostic of MFT dynamo models, the transport 
mechanisms linking the surface field to the dynamo layer must be thoroughly 
understood.

This paper reviews the theory of passive magnetic field transport using mostly 
first (and occasionally higher) order smoothing formalism; the most important 
transport effects are however also independently derived using Lagrangian 
analysis for a simple two-component flow model. Solar 
applications of the theory are also presented. Among some other novel 
findings/propositions it is shown that the observed unsigned magnetic flux 
density in the photosphere requires a small-scale dynamo effect operating
in the convective zone and it is proposed that the net polarity imbalance in
turbulent (and, in particular, hidden) fields may give a major contribution to
the weak large-scale background magnetic fields on the Sun.

\keywords solar physics, magnetism
\end{abstract}
 

\section{Introduction}
\subsection{``Passive'' Fields vs.\ ``Active'' Fields: a Historical Review}
A basic rule of thumb in magnetohydrodynamics (MHD) tells us that the character 
of the interaction between motions and magnetic fields in a (high plasma beta) 
plasma is determined by the ratio of the $E_M$ magnetic and $E_K$ kinetic 
energy densities. If $E_M\ll E_K$ then the Lorentz force may be neglected in 
the equation of motion and our problem is reduced to the \it kinematical case. 
\rm If, on the other hand, $E_M\gg E_K$ then the field will ``channel'' the 
flow and the only potential effect of the motions on the field is the 
generation of small-aplitude MHD waves: this is the \it strong field case. \rm 
Finally, in the \it hydromagnetic case, \rm when $E_M\sim E_K$, there is a 
complicated interaction of flow and magnetic field. Of course we must be aware 
of the fact that $\lang \vc B^2\rang >\lang \vc B\rang^2$ in general, so the 
$E_M$ total magnetic energy density may well exceed the energy density of the 
large-scale mean field. Besides, the validity of the above simple rule may 
possibly also be limited in two dimensions where a weaker magnetic field 
(consisting e.g.\ of a low filling factor set of strong sheets) could possibly 
also influence the motion, owing to the topological constraint (the flow 
cannot ``get around'' the sheets). These latter points were recently brought 
into focus by \cite{Cattaneo+Vainshtein}. Nevertheless, apart from these rather 
obvious caveats, the simple rule summarized above can be considered as correct. 
This simple notion formed the background of MHD thinking in the 1950's and 
60's when mean-field electrodynamics and mean-field MHD were developed 
(\barecite{Parker:cyclonic}, \barecite{Steenbeck+}) for the treatment of the 
kinematic and hydromagnetic case, respectively.

\begin{figure}[htbp]
  \centerline{\psfig{figure=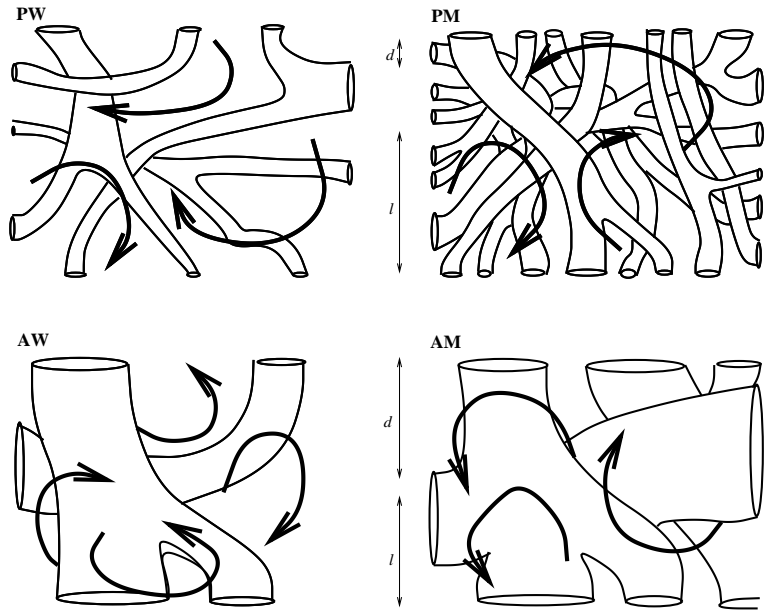,width=10cm,angle=90}}   	
  \caption[]{\label{fig:types}
      Types of magnetic field structure from the point of view of field 
      transport. Note that ``passive'' (P) does not imply ``kinematical'': 
      e.g.\ in case PM the presence of thin flux tubes in large numbers may 
      significantly alter the flow pattern by posing a serious obstacle in the 
      way of the flow.}  
\end{figure}

The picture however got more complicated in the period from the mid-sixties to 
the mid-seventies when solar observations (\barecite{Sheeley:strongfield}, 
\barecite{Stenflo:firstkG}, \barecite{Howard+Stenflo}) and numerical 
experiments (\barecite{Weiss:first}) showed that in the highly conductive 
turbulent solar plasma the magnetic field is concentrated into strong flux 
tubes with very little flux in between. The $B_{\rm t}$ magnetic flux density 
inside the tubes is order of (or greater than) the $B_{\rm eq}$ \it 
equipartition flux density\/ \rm defined by
\[ B_{\rm eq}^2/2\mu =\rho v_{\rm t}^2/2    \]
(SI formula; $\mu$ is the permeability, $\rho$ is the density, $v_{\rm t}=\lang 
v^2\rang^{1/2}$ with $v$ the turbulent velocity). As a consequence of this 
realization, flux tube theory began to develop in the 1970's (see 
\barecite{Parkerbook}, for a review of the results of this period). According 
to flux tube theory, the most important forces acting on a magnetic flux tube 
are the $F_{\rm d}$ aerodynamic drag, the $F_{\rm m}$ magnetic curvature force 
and the $F_{\rm b}$ buoyancy; their approximate expressions are:
\[ F_{\rm b}\sim \frac{B_{\rm t}^2}{2\mu_0 H_P} \qquad F_{\rm m}\sim 
   \frac{B_{\rm t}^2}{\mu_0 R_{\rm c}} \qquad F_{\rm d}\sim\frac{\rho v_{\rm 
   t}^2}d   \]
with $H_P$ the pressure scale height, $d$ the tube diameter and $R_{\rm c}$ the 
curvature radius; in practice, one may put $R_{\rm c}\sim l$ with $l$ the 
characteristic scale of the turbulence. A comparison of these expressions shows 
that for a sufficiently thin flux tube, with a magnetic flux $\Phi <\Phi_{\rm 
cr}=\mbox{min\,}\{l^2, H_P^2\} B_{\rm eq}^4/B_{\rm t}^3$, the drag will 
dominate and the surrounding flow will determine the motion, while thicker 
tubes may move more independently of the surrounding turbulence, under the 
action of dynamical forces. This implies that, for a given energy density 
(for a given $f$ magnetic filling factor or $D$ typical tube separation), 
the transport of the field may be either passive, i.e.\ fully determined by the 
flow between the tubes, or active, i.e.\ to a large degree independent of the 
flow in the non-magnetic component, depending on whether the field is 
organized in a large number of thin fibrils or in a small number of thick flux 
bundles. Our classification of magnetic field types from the point of view of 
transport should therefore be extended into a two-dimensional scheme: 
\smallskip\newline 
\begin{tabular}{l|c|c|}
& Passive & Active \\
& ($d\ll l$) & ($d\sim l$) \\ \hline
Weak ($d\ll D$ or $E_M\ll E_K$) & PW & AW \\ \hline
Moderate ($d\sim D$ or $E_M\sim E_K$) & PM & AM \\ \hline
Strong ($E_M \gg E_K$) & \multicolumn{2}{c|} S  \\ \hline
\end{tabular} \smallskip\newline
(cf.\ also Fig.~\ref{fig:types}). Traditional MFT, with its one-fluid approach, 
may be applied to passive fields, but it is clearly not applicable to the case 
of active fields as these would require a two-fluid description. (An 
interesting attempt at the construction of a more general, two-fluid MFT was 
made by \barecite{Parker:fibrilMFT}.)

Now let us estimate the value of $\Phi_{\rm cr}$ for the solar photosphere. 
Using $H_P\sim 100$\,km, $l\sim 10^3$\,km (the granular scale) and assuming 
$B_{\rm t}\sim B_{\rm eq}\simeq 40$\,mT we find $\Phi_{\rm cr}\sim 
10^9$--$10^{10}$\,Wb. (The assumption of $B_{\rm t}\sim B_{\rm eq}$ is to some 
extent validated by the observations of \barecite{Keller:NATO}.) Flux tubes 
with such low flux values were rarely (if ever) observed on the Sun prior to 
the mid-1980's. As a consequence, all or most of the photospheric magnetic 
field was thought to be in active form, and, from the late 1970's onwards, 
serious doubts arose concerning the relevance of MFT to the solar dynamo 
problem. The emphasis of research shifted to flux tube emergence calculations 
and numerical experiments.

The observations of the last 5--10 years have brought a fundamental change in 
our perception of photospheric fields; an important consequence of these new 
results may be a partial ``rehabilitation'' of MFT as the basic theory of solar 
activity. Because of their importance, these recent observational results will 
be summarized separately in the next subsection. First however we should 
briefly deal with a number of other objections against MFT.

Beginning with the famous paper by \cite{Parker:buoy.prob} on the ``buoyancy 
problem'', evidence has been mounting for a dynamo located at the bottom of the 
solar convective zone or below. It is then clear that the study of magnetic 
field transport is crucial in understanding the cause of this spatial 
restriction of the dynamo mechanism as well as in relating the fields seen at 
the surface to the dynamo deep below. In the past, MFT dynamo models usually 
ignored the transport effects, except turbulent diffusion (and occasionally a 
phenomenological term imitating buoyancy). In fact, although some theoretical 
results concerning magnetic field transport were sporadically published (and 
will be cited in subsequent sections of this paper), this subject has only been 
reviewed once (\barecite{Moffatt:transport}), and solar applications only began 
in earnest in the 1980's.

MFT's usual limitation to the kinematical case with first order smoothing also 
used to be a subject of criticism. In recent years, however, a number of 
important papers on higher order and nonlinear effects (cf.\ Sects.~3.2--3.4) 
improved the position of MFT in this respect as well.

In summary: although MFT is certainly no wonder-drug, it is a powerful and 
versatile theory which is able to reproduce some basic features of solar 
activity (e.g.\ the butterfly diagram) and which should be applicable to at least 
one important component of the solar magnetic field: to passive fields.

\subsection{The Impact of Observations}
Passive fields are freely advected by turbulence, so their fluctuating 
component, the \it turbulent field \/\rm must have a characteristic scale or
correlation length not exceeding that of the turbulence. Numerical experiments 
and turbulence closure models (\barecite{DeYoung}, 
\barecite{Meneguzzi+:smalldyn.simu}, \barecite{Meneguzzi+Pouquet:JFM}) actually 
show that in high Reynolds number non-helical turbulence the maximum of the 
magnetic energy spectrum lies at significantly (about an order of magnitude) 
higher wavenumbers than that of the kinetic energy spectrum. On this basis we 
may expect that the scale of the photospheric turbulent field should be about 
100\,km, well below the resolution limit. How can we hope to observe this 
mixed-polarity field?

The most straightforward approach is to try to increase the resolution of 
magnetograph observations. The presently available resolution of $\sim 
1\arcsec$ is however only sufficient to give us a glimpse of the low-wavenumber 
end of the supposed turbulent magnetic energy spectrum. Observationally, pixels 
with a coherent motion lying far from the network and containing a small 
amount of flux above the noise level may be identified as the resolved 
component of the turbulent field (TF), called intranetwork (IN) field. Any 
component with polarities mixed on a scale smaller than the resolution element 
will remain invisible (hidden field, HF). After tentative detections in the 
1970's (\barecite{Livingston+Harvey}), the systematic study of IN fields began 
in the 1980's (e.g.\ \barecite{Zirin:IN}). The most thorough investigations to 
date were done by Martin (\citeyear{Martin:Kiev}, \citeyear{Martin:NATO}). These 
videomagnetogram studies indicate that the IN elements move coherently from 
pixel to pixel, i.e.\ each of them is the manifestation of a single flux tube 
(with a relatively large flux compared to the other, unseen tubes) rather than 
being a larger scale net polarity fluctuation in the distribution of smaller 
turbulent flux tubes. (This may however partly be a selection effect: the 
magnetograph noise level determination method, as described  e.g.\ by 
\barecite{Murray:polar}, does not distinguish between instrumental and physical 
noise, so the noise-dominated pixels may actually also contain resolved net 
polarity fluctuations in the distribution of the smallest tubes.)  In a typical 
quiet sun area, the unsigned flux density was found to be about 0.5\,mT 
(\barecite{Martin:Kiev}), comparable to the value for ephemeral active regions 
(ER) and for the mixed-polarity network (MN) together, as well as to the cycle- 
and latitude-averaged value for active regions (AR) and unipolar network. 
Noting that the small flux and high inclination of the smallest elements of the 
mixed network (\barecite{Murray:inclin}), which contain most of its unsigned 
flux, indicates that this component of the solar magnetic field should also be 
classified as passive and its fluctuating component may be thought of as part 
of the turbulent field, from this alone we may conclude that turbulent fields 
must give a large fraction of the total unsigned flux density.

But how much unsigned flux resides in the HF? Some indirect methods have been 
devised to study this problem. A rather firm $1\sigma$ upper limit of 10\,mT 
can be placed on $\GB$ on the basis of the lack of any significant Zeeman 
broadening in unpolarized spectral lines (\barecite{Stenflo+Lindegren}). 
Transverse Zeeman linear polarization yields the difference in the field 
components perpendicular to the 
line of sight; again, only upper limits can be given, so the fields are close 
to isotropy (\barecite{Stenflo:lin.pol}). \cite{Stenflo:Hanlelimit} was able to 
place a \it lower\/ \rm limit of 1\,mT on $\GB$ on the basis of an analysis of 
linear polarization measurements, utilizing the Hanle effect. In a more 
detailed analysis of the Hanle effect \cite{Faurob} found that the net unsigned 
flux density in the low photosphere is between 3 and 6\,mT (a recent study, 
\barecite{Faurob+me}, yields a similar but somewhat lower value). This implies 
that the overwhelming majority of $\GB$ is in the form of turbulent fields, in 
line with  a theoretical conclusion based on vertical flux transport models 
(cf.\ Sect.~4.2 below). If the typical scale of the turbulent field is indeed 
not much below 100\,km then the LEST telescope may actually enable observers to 
get a glimpse of a significant fraction of the total unsigned flux.

With this large value of $\GB$, a net surplus of one polarity as little as 1 or 
2 percent over scales exceeding a few times $10^4$\,km will yield a mean \it 
net \/\rm magnetic flux density comparable to the large-scale background 
magnetic field measured outside unipolar network areas. Indeed, the only viable 
alternative to the possibility that this background field is due to the 
turbulent fields is to assume that the surplus magnetic flux in one polarity 
originates from the AR which decay into strong, active network elements \it and 
\/\rm these elements never decay further into small, passive MN/IN/HF elements. 
Before the existence of IN fields and the smallness and high inclination of 
most MN elements were recognized, this assumption was indeed often quoted to 
explain the origin of the large-scale fields. New observations however 
indicate that TF are a more likely candidate to have the polarity surplus. A 
comparison of auto- and cross-correlations of the large-scale polar field by 
\cite{Stenflo:Bab-Leigh} has shown that the rotation law of individual elements 
is quite different from the rotation of the field pattern which must be 
constantly renewed on a timescale between 3 and 30 days; the flux emergence 
rate in AR is not sufficient to replenish the mean field in such a short time. 
Furthermore, high sensitivity magnetograms of the solar disk now show that, in 
contrast to unipolar network areas, the line-of-sight magnetic field in weak 
field areas does not vary too much between center and limb 
(\barecite{Zwaan:Freibg}), i.e.\ these fields have a considerable inclination 
that is difficult to understand if the polarity surplus resides in the strongly 
buoyant thicker network tubes. The weak large-scale field must therefore be the 
result of small net polarity imbalance in the low flux component of the MN 
and/or in the IN/HF, all of which are passive and cannot avoid turbulent 
reprocession into IN fields on a short timescale. (AR however may still be the
ultimate source of this reprocessed flux.) Note that it is even 
possible that most of the large-scale background flux is due to the HF, the 
elements of which are presently unavailable to direct observations!


\section{The main transport mechanisms}
\subsection{Basic Formalism}
Throughout this paper, we will apply the following notations. The letters 
$i$ through $q$ denote integers taking the values 1..$D$ where $D$ is the 
number of dimensions (2 or 3). Other letters stand for reals. Boldface and 
overarrow denotes 
vectors, hat denotes second-order tensors. For repeated small integer italic 
indices summation is implied---but not for capital indices! 
$d_x=d/dx$; $\ptl_x=\ptl /\ptl x$; $\ptl_i=\ptl/\ptl x_i$;
$\delta_{ij}$ is the unit tensor, $\epsi_{ijk}$ is the Levi-Civita tensor. 

We start from the first two Maxwell equations supplemented with Ohm's law:
\[ \nabla\times\vc B =\mu\vc j \qquad \nabla\times\vc E =-\ptl_t\vc B \]
\[ \vc j =\mx\sigma (\vc E+\vc u\times\vc B) , \]
with $t$ the time, $\vc u$ the velocity and $\mx\sigma$ the (tensorial) conductivity.
From this
\[ \ptl_t\vc B=\nabla\times (\vc u\times\vc B) -\nabla\times\left[ (\mx\eta 
   \nabla)\times\vc B\right]  ,  \]
or in components:
\[ \ptl_t B_i =\epsi_{ijk}\ptl_j\epsi_{klm}u_lB_k -\epsi_{ijk}\ptl_j 
   \epsi_{klm}\eta_{lp}\ptl_p B_m  .  \label{eq:anitransp} \]
In a gas, the $\mx\eta=(\mu\mx\sigma )^{-1}$ turbulent magnetic diffusivity is 
isotropic, so $\eta_{kl}=\eta\delta_{kl}$ and the induction equation finally 
takes the form
\[ \ptl_t\vc B=\nabla\times (\vc u\times\vc B) -\nabla\times\eta 
   (\nabla\times\vc B ) . \label{eq:induction} \]
In a turbulent medium both $\vc u$ and $\vc B$ may be split into average and 
fluctuating parts:
\[ \vc b =\lang\vc B\rang + \vc B' \qquad \vc u =\vc U +\vc v . \]
(We take ensemble averages, ignoring here the problem of relating these 
expected values to actual measurements. The Reynolds rule $\lang\vc B'\rang =
\lang\vc v\rang=0$ is assumed.) Our aim is now to derive the evolution equation 
analogous to (\ref{eq:induction}) for the $\lang\vc B\rang$ mean field.
The average of (\ref{eq:induction}) is:
\[ \ptl_t\NB =\nabla\times(\vc U\times\NB +\vec{\cal E}) 
   -\nabla\times\eta\nabla\times\langle\vc B\rangle  .   \label{eq:avind} \]
The real task then consists in finding an expression for the 
\[ \vec{\cal E} = \lang\vc v\times\vc B'\rang   \]
turbulent electromotive force.

In a discussion of magnetic field transport, after the derivation of the 
equation for $\NB$ we will face the problem of distinguishing the terms 
responsible for transport effects from other terms. To some degree this is a 
matter of definition: the only reason why we first wrote down the anisotropic 
form (\ref{eq:anitransp}) of the induction equation is that 
(\ref{eq:anitransp}) is the \it par excellence \/\rm  transport equation for 
an advected solenoidal vector field---its first term expressing advection with 
the flow, its second term describing an anisotropic diffusion. One is tempted 
to say that those terms of (\ref{eq:avind}) that are formally identical to the 
terms of (\ref{eq:anitransp}) should be referred to as ``transport terms''; 
however, from a conceptual point of view, our choice of definition should be 
such that the appearance of transport effects and of non-transport effects, 
respectively, will not 
depend on the chosen reference frame---i.e.\ only full tensorial terms obeying 
tensor transformation rules should be referred to as ``transport terms''. Our 
approach will be to define ``transport effects'' as described by those 
and only those tensorial terms of (\ref{eq:avind}) (or of $\vec{\cal E}$) that
in some reference frame(s) give rise to terms formally identical to those of 
(\ref{eq:induction}).

In order to get an expression for $\vec{\cal E}$, we subtract (\ref{eq:avind}) 
from (\ref{eq:induction}) to get 
\[ \ptl_t\vc B' = \nabla\times(\vc U\times\vc B' +\vc v\times\NB +\vc G) 
    -\nabla\times\eta\nabla\times\vc B' , \label{eq:fluct} \]
\[ \vc G =\vc v\times\vc B' -\vec{\cal E} .    \]
Now, assuming that the fluctuating magnetic (and velocity) fields have a 
correlation time $\tau$ (independent of location) over which all memory is 
lost,  we may write
\[ \vec{\cal E} =\lang\vc v\times\int_0^\tau \ptl_t\vc B'\,dt\rang\sim \lang\vc v 
   \times\ptl_t\vc B'\rang\tau  . \label{eq:emfexpr} \]
Substituting here (\ref{eq:fluct}), we get an approximate expression for 
$\vec{\cal E}$. In principle, a more exact formula might be derived by leaving the 
integral in (\ref{eq:emfexpr}) instead of using multiplication of the integrand 
with $\tau$ as a crude estimate. This would result in integrals involving 
Green's functions (\barecite{Radler:IAUSymp}). However, for a practical 
application of the method an order-of-magnitude estimate must be introduced 
again at the end, as the correlations appearing in the integrands are not 
known. The Fourier transform method (\barecite{Moffatt:transport}) encounters a 
similar problem (with the added difficulty of treating inhomogeneity in a 
Fourier representation). It will spare us a lot of work to make this 
unavoidable approximation right here at the outset. 

After the substitution, a third-order correlation will appear in the expression 
for $\vec{\cal E}$ (itself a second-order correlation) owing to the quantity $\vc 
G$. A similar expression for $\vc G$, in turn, would involve a fourth-order 
correlation, etc., so an infinite hierarchy of moment equations appears---a 
problem familiar from turbulence theory. In order to make the problem 
tractable, this chain must be broken somewhere. This is usually achieved 
by the so-called \it smoothing, \rm i.e.\ by assuming that all correlations 
higher than a certain order disappear. (An alternative method will be applied 
in section 2.4). In the simplest case of first-order 
smoothing (also known as \it second order correlation approximation \/\rm  or 
SOCA) one assumes $\vc G=0$. A sufficient (but not necessary!) condition for 
this is that $B'\ll |\NB|$, which is true if the Strouhal number 
$v_{\rm t}\tau/l\ll 1$ \it and \/\rm $v_{\rm t}\tau/H_B\ll 1$, $H_B$ being the 
scale length of $\NB$. 
(Another sufficient condition, of less importance 
in the solar context, is that the magnetic Reynolds number 
$lv_{\rm t}/\eta\ll 1$.) The expression for $\vec{\cal E}$ 
will be derived in Sect.~2.3. First, 
however, we will discuss the general form of $\vec{\cal E}$ and the form of the 
transport equations.

\subsection{Symmetry Considerations and Double-Scale Analysis}
Contrary to most discussions of MFT, this paper concentrates on the transport 
effects and does not deal with the dynamo effects. (For a review of the MFT of 
the solar dynamo the reader is referred to the papers by \barecite{Radler80} 
and 
\barecite{Hoyng:NATO}.) Dynamo terms may be defined as those tensorial terms 
that would lead to an increase of the field without limit (in the kinematic 
description), were there no other terms present. 
Unfortunately, transport terms and dynamo terms do not form two disjunct sets, 
so the two problems cannot be discussed independently in the general case. In 
order to avoid any disturbing dynamo effect in our ``sterile'' study of the 
transport, we assume that all dynamo terms vanish. A (strictly unproven but 
very likely) conjecture states that a sufficient condition for this is that the 
velocity field should be \it parity invariant\/ \rm  in the sense that all its 
mean pseudoscalar quantities vanish. From this point on, we therefore restrict our 
discussion to the case of parity invariant velocity fields.

For a discussion of what this implies for the flow, we introduce the term 
\it preferred orientation. \rm If a second order tensor with a non-degenerate 
eigenvalue can be constructed from the velocity field, then we say that the 
corresponding set of eigenvectors defines a preferred orientation. Furthermore, 
the flow will be said to have a \it preferred sense of motion, \rm if a
non-zero polar vector can  be constructed from the velocity field. 
In the same spirit, a \it preferred sense of rotation \/\rm is said to exist 
if a non-zero pseudo-vector functional $\vc f[\vc u(\vc x)]$ 
can be constructed \it and\/ \rm if the flow is \it 
either\/ \rm not parity invariant \it or\/ \rm $\vc f$ 
has the property $\partial_i f_i=0$. A \it preferred direction\/ \rm 
implies either a preferred sense of motion or a preferred sense of rotation. 
A preferred direction also defines a preferred orientation---but the 
reverse does not hold.

The $\vc u(\vc x)$  velocity field must obey the equation of motion; from the 
structure of the equation of motion it follows that in the kinematical case any 
deviation from statistical isotropy must be the consequence of either a 
non-vanishing $\vc g$ gravity or of a non-vanishing $\vec\omega$ 
large scale vorticity of the flow. The assumption of parity invariance then 
requires $\vc g\vec\omega =0$. This shows that we must either limit our study 
to the equator, or ignore stratification, or ignore rotation. It is this last 
assumption, i.e.\ the neglect of large-scale rotation that seems to be the 
least restrictive in the solar case, as it still allows us to incorporate 
some indirect effects of rotation in our model, notably
\begin{enumerate}
\item A deviation from spherical symmetry (but not from axial symmetry). In a 
spherical Sun with $\theta\phi r$ polar coordinates we 
therefore allow $\partial_\theta\neq 0$ for mean quantities, but we require 
$\lang\partial_\phi\rang=0$. In this sense our discussion will be limited to 
2-dimensional models (although the fluctuations may be 3D). As a particularly 
simple case, we will also regard plane parallel geometry with  the $z$-axis 
parallel to $\vc g$ and $\lang\partial_y\rang=0$; this can be viewed as a local 
approximation to the spherical case (corresponding to $r\rightarrow\infty$).
\item The existence of other preferred senses of motion beside $\vc g$. The 
parity invariance condition however implies that all such senses of motion must 
lie in one plane, which is why we had to assume axial symmetry. 
(Otherwise, with $\vc f$, $\vc g$, $\vc h$ being 3 preferred 
polar vectors, the pseudoscalar $(\vc f\times\vc g)\vc h$ would not vanish.) 
Any preferred sense of rotation must then be perpendicular to this plane.
The plane will be the meridional plane. Then it follows that for any $\mx a$ 
polar tensor we have $a_{\phi r}=a_{r\phi}=a_{\theta\phi}=a_{\phi\theta}=0$, 
while for pseudo-tensors $a_{\theta\theta}=a_{rr}=a_{\theta r}=a_{r\theta}=0$.
(Otherwise, $\mx a\vc g$ would define a polar vector outside the meridional 
plane or a pseudo-vector in the meridional plane, respectively.) 
\item In principle, a meridional circulation can also be included, as its 
large-scale vorticity would lie in the azimuthal direction where a preferred 
sense of rotation is not excluded. In the present discussion, however, for 
simplicity we assume $\vc U=0$, i.e.\ we neglect both rotation and circulation.
\end{enumerate}

Now if we assume that the scale $H_B$ of the large-scale mean field is (much) 
larger than $v_{\rm t}\tau$ then, within a correlation time $\tau$, mean 
quantities like $\vec{\cal E}$ will only ``feel'' the large-scale field within 
a radius small compared to $H_B$, so we can expand $\vec{\cal E}$ in terms of 
$\NB$ as
\[ {\cal E}_i=-\alpha^{(T)}_{ik}\lang B_k\rang 
   -\beta^{(T)}_{ijk}\partial_j \lang B_k\rang   . \label{eq:2scale}  \]
(The minus sign is just a convention.)
From (\ref{eq:emfexpr}) and (\ref{eq:fluct}) it follows that each term of 
$\vec{\cal E}$ must include the Levi-Civita tensor as one of its factors. 
Therefore we may write
\[ \beta^{(T)}_{ijk}=\epsi_{ijl}\beta^{(0)}_{lk}
  +\epsi_{ilk}\beta^{(T)}_{lj}
   +\beta^{(r)}_{ijk} , \]
where $\beta^{(0)}_{lk}$ and $\beta^{(r)}_{ijk}$ will certainly vanish in 
SOCA. (This is because 
``there are not the proper number of indices'' in the first moment 
of (\ref{eq:fluct}). 
For the same reason, in SOCA we cannot expect higher order terms in the 
expansion (\ref{eq:2scale}).) We now split 
$\mx\alpha^{(T)}$ and $\mx\beta^{(T)}$ into symmetric and antisymmetric parts, 
further splitting the symmetric part of $\mx\alpha^{(T)}$ into an isotropic and 
a ``traceless'' part:
\[ \alpha^{(T)}_{ik}=-\alpha\delta_{ik}+\tilde{\alpha}_{ik}
   +\epsi_{ijk}\gamma_j   \label{eq:alphexpr} \]
\[ \beta^{(T)}_{ik}=\beta_{ik}+\epsi_{ijk}\delta_j   \]
\[ \tilde\alpha_{ii}=0 \qquad \tilde\alpha_{ij}=\tilde\alpha_{ji} \qquad 
   \beta_{ij}=\beta_{ji} \]
With this, the general expression of the turbulent electromotive force 
(neglecting the higher order $\beta^{(r)}_{ijk}$ term) is
\[ \vec{\cal E}=\alpha\langle\vc B\rangle
   -\mx{\tilde\alpha}\langle\vc B \rangle-\vec\gamma\times\langle\vc B\rangle 
   -(\mx\beta\nabla)\times\langle\vc B\rangle
   -(\vec\delta\times\nabla)\times\langle\vc B\rangle  , \]
or
\[ {\cal E}_i=\alpha\delta_{ik}\lang B_k\rang -\tilde\alpha_{ik}\lang B_k\rang  
   -\epsi_{ijk}\gamma_j \lang B_k\rang 
   -\epsi_{ijk}\beta_{jl}\partial_l\lang B_k\rang 
   -\delta_k\partial_i\lang B_k\rang .   \]
(Note that the brackets are redundant and they only serve to emphasize that 
our definition of $\mx\beta$ and $\vec\delta$ differs from that used in some 
other papers.)

Writing this into (\ref{eq:avind}) we see that the role of $-\vec\gamma$ is 
similar to that of $\vc U$ i.e.\ this term describes the advection of the mean 
field with an effective velocity $-\vec\gamma$ without real large-scale 
motions. This effect is called \it 
(normal) pumping \/\rm of the magnetic field. $\vec\gamma$ being a polar 
vector, the pumping requires the existence of a preferred sense of motion of 
the field. Such a preferred sense of motion may e.g.\ be due to a gradient in 
$\rho$ (\it density pumping\/\rm ), to a gradient in $v_{\rm t}$ (\it turbulent 
pumping\/\rm ), to  a topological asymmetry of the flow (\it topological 
pumping\/\rm ), etc. 

The role of the $\mx\beta$ term is analoguous to that of the 
diffusive term in (\ref{eq:avind}). The $\mx\beta$-term however does not only 
describe an anisotropic diffusion, but its form is even more general than the 
corresponding term in (\ref{eq:anitransp}), including a ``diffusion'' of the 
magnetic field along its own direction. For this reason this term is sometimes 
split in two parts corresponding to the symmetric and antisymmetric parts of 
the $\partial_i B_k$ tensor (e.g.\ \barecite{Radler80}); this is  
however just a formal manipulation as the coefficients of these terms will not 
be independent. 

In a parity invariant flow the pseudoscalar $\alpha$ will vanish. But what 
about $\mx{\tilde\alpha}$? Writing down the components of $\vec{\cal E}$ one 
finds that the role of this term shows some similarity to that of the 
$\vec\gamma$-term: it apparently describes a pumping---with a sign that depends 
on the orientation of the magnetic field! If the direction of this \it 
anomalous pumping \/\rm is characterized by a $\vec{\tilde\gamma}$ vector 
then there 
exists a $\vec\lambda$ vector perpendicular to $\vec{\tilde\gamma}$ such that 
the field component parallel to $\vec\lambda$ will be transported along 
$\vec{\tilde\gamma}$, while the component perpendicular to $\vec\lambda$ (and 
to $\vec{\tilde\gamma}$) will be transported in the opposite direction. 
(In a similar way, the $\vec\delta$-term leads to an anomalous or ``skewed'' 
diffusion effect.) 
As all 
terms of $\vec{\cal E}$ must include the Levi-Civita tensor, the general form 
of $\mx{\tilde\alpha}$ is
\[ \tilde\alpha_{ik}=\epsi_{ijl}\tilde\gamma_j C_{lk}  , \]
to be compared with the $\gamma$-term in (\ref{eq:alphexpr}). The dependence on 
$\vec\lambda$ must obviously be contained in the polar tensor $\mx C$. Its 
general form is
\[ C_{lk}=\delta_{lk}+b\lambda_l\lambda_k+c\epsi_{klm}\lambda_m  .  \]
(A free factor before the first term was omitted, as it may be defined into 
$\vec{\tilde\gamma}$.) As in the kinematical description the pumping may only 
depend on the orientation of the magnetic field, but not on its direction, $\mx 
C$ may only involve terms of even order in $\lambda_i$, i.e.\ $c=0$. 
The symmetry condition on $\mx C$ then implies $b=-2$. 

Summarizing these considerations, we find that the general SOCA expression of 
$\vec{\cal E}$ for a parity invariant flow  must have the form
\[ {\cal E}_i= 
   -\epsi_{ijl}\tilde\gamma_j(\delta_{lk}-2\lambda_l\lambda_k)\lang B_k\rang 
   -\epsi_{ijk}\gamma_j \lang B_k\rang 
   -\epsi_{ijk}\beta_{jl}\partial_l\lang B_k\rang 
   -\delta_k\partial_i\lang B_k\rang .   \label{eq:finalE} \]
Substituting this into (\ref{eq:avind}) we find the general form of the mean 
field transport equation in an axially symmetric geometry with 
$\lambda_i=\delta_{i2}$, $\vc U=0$ (and therefore $\vec\delta=0$):
\[ \partial_t A_\phi=a_{\theta\theta}\partial^2_\theta A_\phi +a_{\theta r}
   \partial_\theta\partial_r A_\phi + a_{rr} 
   \partial^2_r A_\phi +a_\theta\partial_\theta A_\phi +a_r\partial_r A_\phi 
   +a_0 A_\phi  \label{eq:poltransp}  \label{eq:spheri} \]
\[ \lang B_\theta\rang =-A_\phi /r-\partial_r A_\phi \qquad 
   \lang B_r\rang =\partial_\theta A_\phi /r +\cot\theta A_\phi /r   \]
\[ \partial_t \lang B_\phi\rang =b_{\theta\theta}\partial^2_\theta 
   \lang B_\phi\rang  +b_{\theta r}
   \partial_\theta\partial_r \lang B_\phi\rang  +b_{rr} 
   \partial^2_r \lang B_\phi\rang  +b_\theta\partial_\theta \lang B_\phi\rang  
   +b_r\partial_r \lang B_\phi\rang +b_0 \lang B_\phi\rang 
    \label{eq:tortransp}  \label{eq:spheritor}  \]
where
\[ a_{\theta\theta}=b_{\theta\theta}= \beta_{\theta\theta}/r^2 \qquad 
   a_{\theta r}=b_{\theta r}= 2\beta_{\theta r}/r 
   \qquad a_{rr}=b_{rr}=\beta_{rr} \]
\[ a_\theta = (\gamma_\theta +\tilde\gamma_\theta )/r 
   +(\beta_{\theta\theta }\cot\theta -\beta_{\theta r})/r^2    \]
\[ a_r =(\gamma_r+\tilde\gamma_r) +(\beta_{rr}
  +\beta_{\theta\theta} +\beta_{\theta r}\cot\theta )/r \]
\[ a_0=[(\gamma_r +\tilde\gamma_r) +\cot\theta (\gamma_\theta 
  +\tilde\gamma_\theta )]/r-[\beta_{rr}+\beta_{\theta\theta}\cot^2\theta 
  +\beta_{\theta r}(1+\cot\theta )]/r^2  \]
\[ b_\theta = (\gamma_\theta -\tilde\gamma_\theta )/r 
   +\beta_{\phi\phi}\cot\theta /r^2 +\partial_r\beta_{\theta r}/r 
   +\partial_\theta\beta_{\theta\theta}/r^2   \]
\[ b_r =(\gamma_r -\tilde\gamma_r) +(\beta_{rr}
   +\beta_{\phi\phi})/r +\partial_r \beta_{rr} +\partial_\theta\beta_{\theta 
   r}/r \]
\begin{eqnarray}
   b_0= & [(\gamma_r -\tilde\gamma_r) +\partial_\theta 
   (\gamma_\theta -\tilde\gamma_\theta )+\partial_r\beta_{\phi\phi}]/r 
   +\partial_r(\gamma_r -\tilde\gamma_r ) \\
   & +[\cot\theta\,\partial_\theta\beta_{\phi\phi} 
   -\beta_{\phi\phi}/\sin^2\theta)]/r^2   
\end{eqnarray}
The molecular diffusivity term was ignored, as it can always be defined into 
the $\beta$-term and is usually negligible. Note that for non-parity-invariant 
turbulence the appearance of $\beta_{\theta\phi}$ and $\beta_{\phi r}$ would 
give rise to dynamo terms, i.e.\ the corresponding part of the 
$\mx\beta$-term is a transport \it and\/ 
\rm a dynamo term at the same time, illustrating our above point on the 
impossibility of a clean distinction between dynamo and transport effects. (The 
dynamo effect in question was discovered by \barecite{Radler80}, who mentioned 
it in the context of a ``$\beta\omega$-dynamo''.)

In the plane parallel case these equations reduce to the particularly simple 
form
\[ \partial_t A_y= \beta_{xx}\partial^2_x A_y +2\beta_{xz}\partial_x \partial_z 
   A_y +\beta_{zz} \partial^2_z A_y +(\gamma_x +\tilde\gamma_x )\partial_x A_y 
   +(\gamma_z +\tilde\gamma_z )\partial_z A_y \label{eq:plantransp}   \]
\[ \lang B_x\rang =-\partial_z A_y \qquad \lang B_z\rang =\partial_x A_y 
   \label{eq:planpol} \]
\begin{eqnarray} 
   \partial_t\lang B_y\rang = & \beta_{xx}\partial^2_x \lang B_y\rang  
   +2\beta_{xz}\partial_x \partial_z \lang B_y\rang  +\beta_{zz} 
   \partial^2_z \lang B_y\rang +(\partial_x\beta_{xx}+\partial_z\beta_{xz})
   \partial_x\lang B_y\rang \nonumber \\
   & +(\partial_z\beta_{zz}+\partial_x\beta_{xz})
   \partial_z\lang B_y\rang +[\partial_z(\gamma_z-\tilde\gamma_z) 
   +\partial_x(\gamma_x-\tilde\gamma_x)]\lang B_y\rang  
\end{eqnarray}

\subsection{First Order Smoothing}
It remains to find the expressions of the coefficients appearing in the 
transport 
equations. For this purpose we substitute (\ref{eq:fluct}) in 
(\ref{eq:emfexpr}) under the assumption $\vc U=\vc G=0$ and find
\begin{eqnarray} 
   {\cal E}_i= & -\tau\epsi_{ikl}\lang v_j v_k\rang\partial_j \lang B_l\rang 
   -\tau \epsi_{ikl}\partial_j\lang v_j v_k\rang \lang B_l\rang \nonumber \\
   & +\tau\epsi_{ikl}\lang 
   v_k\partial_j v_j\rang \lang B_l\rang -\tau\epsi_{jkl}\lang v_k\partial_i 
   v_j\rang \lang B_l\rang     \label{eq:4terms} 
\end{eqnarray}
Comparing this with (\ref{eq:finalE}) and assuming that the fluid obeys the
$\ptl_i(\rho u_i)=0$ anelastic continuity equation (a good approximation in the
solar convective zone), taking $\lambda_i =\delta_{iQ}$ and 
$\epsi_{IKQ}\neq 0$ we arrive at
\[ \beta_{ij}=\tau\lang v_i v_j\rang \qquad \delta_i=0 \label{eq:betakif} \]
\[ \gamma_k=\tau\partial_j\lang v_j v_k\rang /2 \label{eq:gammakif} \]
\[ \tilde\gamma_K = \tau\tfrac{{}_K{\cal A}^I_Q-1}{{}_K{\cal A}^I_Q+1}
   (\lang v_K v_i \partial_i\rho/\rho +\partial_K\lang v^2_K\rang /2) 
   +(\partial_I\lang v_I v_K\rang 
   -\partial_Q\lang v_Q v_K\rang )/2    \label{eq:tildegammakif} \]   
\[ {}_K{\cal A}^I_Q =\frac{\lang v_K\partial_I v_I\rang}{\lang v_K\partial_Q 
   v_Q\rang} .   \]
We see that the normal pumping only contains a term proportional to the 
gradient of the turbulent velocity correlation tensor, i.e.\ it is a pure 
turbulent pumping. The anomalous pumping, in contrast, also contains terms 
related to the density gradient. 

In the transport equations  $\gamma_i$ and $\tilde\gamma_i$ only occur in 
combinations $\gamma_i\pm\tilde\gamma_i$. In spherical coordinates these 
combinations are:
\begin{eqnarray}
   \gamma_\theta +\tilde\gamma_\theta = & 
   \tau\tfrac{{}_1{\cal A}^3_2-1}{{}_1{\cal 
   A}^3_2+1}\lang v_\theta v_r\rang d_r\rho /\rho +\tau\tfrac{{}_1{\cal 
   A}^3_2}{{}_1{\cal A}^3_2+1}(\partial_\theta\lang v^2_\theta\rang 
   +2\lang v_\theta v_r\rang )/r     \nonumber \\ 
   & +\tau\partial_r\lang v_\theta v_r\rang  
\end{eqnarray}
\begin{eqnarray}
   \gamma_r +\tilde\gamma_r = & \tau\tfrac{{}_3{\cal A}^1_2-1}{{}_3{\cal 
   A}^1_2+1}\lang v^2_r\rang d_r\rho /\rho +\tau\tfrac{{}_3{\cal 
   A}^1_2}{{}_3{\cal A}^1_2+1}\partial_r\lang v^2_r\rang \nonumber \\
   & +\tau\partial_\theta\lang v_\theta v_r\rang /r +\tau(\lang v^2_r\rang -\lang 
   v^2_\theta\rang )/r  
\end{eqnarray}
\begin{eqnarray}
   \gamma_\theta -\tilde\gamma_\theta = & 
   -\tau\tfrac{{}_1{\cal A}^3_2-1}{{}_1{\cal 
   A}^3_2+1}\lang v_\theta v_r\rang d_r\rho /\rho +\tau\tfrac{1}{{}_1{\cal 
   A}^3_2+1}(\partial_\theta\lang v^2_\theta\rang + 2\lang v_\theta 
   v_r\rang )/r \nonumber \\
   & +\tau\lang v_\theta v_r\rang /r +\tau\cot\theta (\lang 
   v^2_\theta\rang -\lang v^2_\phi\rang )/r  
\end{eqnarray}
\begin{eqnarray}
 \gamma_r -\tilde\gamma_r = & -\tau\tfrac{{}_3{\cal A}^1_2-1}{{}_3{\cal 
   A}^1_2+1}\lang v^2_r\rang d_r\rho /\rho +\tau\tfrac{1}{{}_3{\cal 
   A}^1_2+1}\partial_r\lang v^2_r\rang \nonumber \\
   & +\tau\cot\theta\lang v_\theta v_r\rang /r +\tau(\lang v^2_r\rang -\lang 
   v^2_\phi\rang )/r  
\end{eqnarray}
In the plane parallel case we have
\[ \gamma_x +\tilde\gamma_x =\tau\tfrac{{}_1{\cal A}^3_2-1}{{}_1{\cal 
   A}^3_2+1}\lang v_x v_z\rang d_z\rho /\rho +\tau\tfrac{{}_1{\cal 
   A}^3_2}{{}_1{\cal A}^3_2+1}\partial_x\lang v^2_x\rang 
     +\tau\partial_z\lang v_x v_z\rang  \] 
\[ \gamma_z +\tilde\gamma_z =\tau\tfrac{{}_3{\cal A}^1_2-1}{{}_3{\cal 
   A}^1_2+1}\lang v^2_z\rang d_z\rho /\rho +\tau\tfrac{{}_3{\cal 
   A}^1_2}{{}_3{\cal A}^1_2+1}\partial_z\lang v^2_z\rang 
   +\tau\partial_x\lang v_x v_z\rang \]
\[ \gamma_x -\tilde\gamma_x =-\tau\tfrac{{}_1{\cal A}^3_2-1}{{}_1{\cal 
   A}^3_2+1}\lang v_x v_z\rang d_z\rho /\rho +\tau\tfrac{1}{{}_1{\cal 
   A}^3_2+1}\partial_x\lang v^2_x\rang  \]
\[ \gamma_z -\tilde\gamma_z =-\tau\tfrac{{}_3{\cal A}^1_2-1}{{}_3{\cal 
   A}^1_2+1}\lang v^2_z\rang d_z\rho /\rho +\tau\tfrac{1}{{}_3{\cal 
   A}^1_2+1}\partial_z\lang v^2_z\rang   \]

For horizontally isotropic turbulence the vertical component of the anomalous 
pumping clearly vanishes 
(as we can expect from symmetry considerations). For a horizontal anisotropy 
induced by rotational influence we expect $\lang v^2_\theta\rang <\lang 
v^2_\phi\rang$, ${}_1{\cal A}^3_2 \la 2$, ${}_3{\cal A}^1_2 <1$ and $\lang v_r 
v_\theta\rang <0$; in this case, the density pumping (i.e.\ the first term in 
each of the above expressions, proportional to $\nabla\rho$) is directed 
downwards and 
polewards for poloidal fields and in the reverse direction for toroidal fields. 
On the other hand, the anomalous part of the turbulent pumping (to which the 
main contribution is given by the second 
term) acts in the opposite direction, if the turbulent velocity decreases 
inwards (in the bulk of the solar convective zone). In reality, all the 
anisotropy parameters involved should be uniquely determined by the 
$\Omega$ rotational velocity and by the form of the velocity covariance tensor 
for the ``original turbulence'' (i.e.\ for the non-rotating case). 
E.g.\ assuming a quasi-isotropic original turbulence, \cite{Kichat:dens.pump} 
found that the density pumping is always perpendicular to the rotational axis. 


\subsection{Lagrangian Analysis for a Two-Component Flow}
Despite its generality and elegance, the first (and higher) order smoothing 
discussion of the magnetic field transport suffers from two major drawbacks. 
First, the conditions of its applicability are unclear ($B^\prime\ll |\NB|$ and 
a small Strouhal number are just sufficient conditions); second, the 
transport effects found by this method do not readily lend themselves to a 
simple physical interpretation. An alternative method of finding an approximate 
expression for the turbulent electromotive force is the so-called Lagrangian 
analysis; in essence, this involves the neglect of the diffusive term in 
(\ref{eq:fluct}), in contrast to SOCA, where the $\vc G$-term was neglected. 
In the present paper we do not present a general, abstract Lagrangian 
analysis of the magnetic field transport problem (see 
\barecite{Moffatt:transport} for such a discussion in the incompressible case). 
Instead, we limit ourselves to the case of a very simple (but anelastic, not 
incompressible) plane parallel 2D flow. Because of its simplicity, our 
description will be 
able to give an easy-to-grasp physical picture of the transport.

Our 2D flow is assumed to be periodic in the horizontal ($x$-) direction. Of 
each period of length $2l$, a fraction $f_u$ is occupied by a local upflow, and 
a fraction $f_d$ is occupied by a downflow ($f_u+f_d=1$). This flow pattern is 
persistent for a time $\tau$, then new up- and downflows will instantaneously 
form with a random phase shift compared to the old flow. At the same time, 
diffusion is assumed to smooth the large scale field, destroying any 
correlations with the flow. (See \barecite{Moffatt:transport}, for the general 
background of these assumptions.)
The vertical ($z$-) velocity is assumed to be horizontally 
constant in both the up- and downflows at a value of $v_u$ and $v_d$, 
respectively. The jump at the up/downflow boundary is of course unphysical, but 
the present model only serves to give us a clue to the essence of the transport 
effects. We introduce a large-scale horizontal magnetic field which may lie in 
the plane of the motions or perpendicular to it. The layer may be stratified in 
$\rho$, $v=(v_u-v_d)/2$ and $\NB$. For simplicity, we only regard 
\it one\/ \rm type of inhomogeneity at a time. The horizontal velocity is found 
from the anelastic continuity equation for each of these cases. We further 
assume 
\[ l\sim v\tau\ll H  \label{eq:assump} \] 
where $H$ is the scale height of the inhomogeneous quantity 
on which the transport depends. Then we have $\lang v^2_x\rang\ll v^2$, so the 
kinetical energy flux may be written as
\[ 2 F_{\rm kin}/\rho =f_uv_u^3+f_dv_d^3 , \label{eq:2compFkin} \] 
while the anelastic continuity equation implies
\[ f_u v_u+f_d v_d=0 .  \label{eq:2compcont} \]
In the case when there is no inhomogeneity in $v$, we expect $F_{\rm kin}=0$, 
and the above relations imply $f_u=f_d=0.5$. 

\begin{figure}[htbp]
  \centerline{\psfig{figure=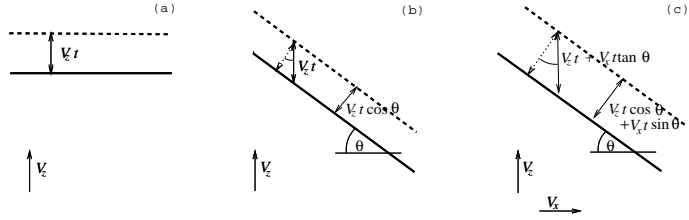,width=9cm,angle=90}}   	
  \caption[]{\label{fig:tiltedtube} The tilt of a flux tube does not influence 
  its vertical transport in a purely vertical flow, but a horizontal flow will 
  give it an extra lift.}  
\end{figure} 

Before we discuss the individual transport effects, some general remarks have 
to be made. The first point is illustrated in Fig.~\ref{fig:tiltedtube}. In a 
purely vertical flow, the vertical transport of the field does not depend on 
its angle to the horizontal: this shows that our results for a horizontal 
large-scale field will remain valid for a tilted field as long as no 
large-scale horizontal flows exist. A horizontal flow, on the other hand, will 
give a vertical lift to a tilted tube---this effect will be important in 
Sect.~2.6 below.

\begin{figure}[htbp]
  \centerline{\psfig{figure=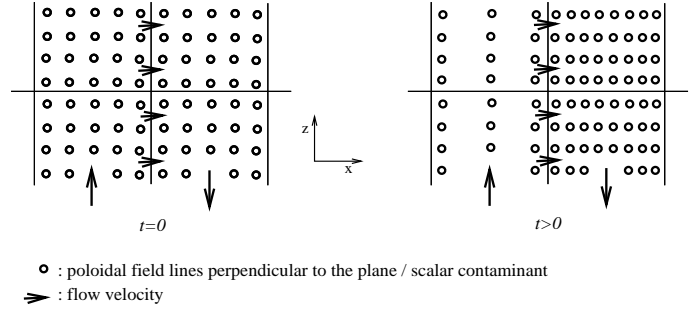,width=9cm,angle=90}}   	
  \caption[]{\label{fig:polpump} Density pumping for a poloidal field.}  
\end{figure}

Another remark concerns the condition (\ref{eq:assump}). We will find that 
under this assumption the SOCA expressions for the field transport effects are 
returned. In the Sun $H\geq l$ for any mean quantity (cf.\ Sect.~4), so this 
condition is certainly less restrictive than that of the small Strouhal number. 
Furthermore, the assumption $B^\prime\ll |\NB |$ is not needed. In fact, even 
in SOCA it would have been sufficient to assume that it is only the part of the 
fluctuating field \it correlated with \/\rm the turbulent velocity that is 
small, while large fluctuations of independent origin (e.g.\ due to turbulent 
concentration by small-scale turbulence) may exist. In other words, the thick 
lines in Figs.~\ref{fig:polpump}--\ref{fig:turpump} may represent field 
lines of a large-scale diffuse field \it or \/\rm thin flux tubes embedded in a 
non-magnetic plasma. In this latter case, the direction of the individual tubes 
is clearly also irrelevant for the transport, which implies that \it each of 
the transport terms appearing in the transport equation of $\NB$ must also 
appear in the transport equation of $\GB$, \rm assuming that $\GB$ is mostly 
due to a 
strongly intermittent, connected field structure, and not to a number of 
isolated small flux rings. The reverse of this statement is however not 
true, as we will see in Sect.~3.2 below.

\subsection{Turbulent Diffusion}
For the study of turbulent diffusion we assume $\nabla\rho=\nabla v=0$, while 
the large scale horizontal magnetic field may depend on $z$. With $v_x\equiv 0$ 
and letting $B_0(z)=B_x(z;t=0)$ we have $B_x(x,z;\tau)=B_0[z-v_z(x)\tau]$. 
Then, assuming $v\tau\ll H_B$, we have
\[ {\cal E}_y =\lang v_z B_x\rang =-\tau v^2 d_z B_0 , \]
independently of the orientation of the field.
The form of the coefficient, $\tau v^2$ is to be compared with the SOCA 
expression (\ref{eq:betakif}). The physical interpretation is of course 
straightforward: if e.g.\ the field increases downwards, upflows bring more 
flux than downflows, resulting in a net transport upwards. The use of a 
turbulent diffusivity or conductivity was first proposed by 
\cite{Sweet:turb.cond} and \cite{Csada}.

\subsection{Density Pumping}
The only inhomogenous mean quantity is now $\rho$, increasing downwards. 
We study a single half-cell $[-0.5<x/l<0.5]$, representing the whole 
pattern; $x=0$ is fixed at an up/downflow boundary with $x>0$ corresponding to
a downflow. The horizontal flow is then 
\begin{eqnarray}
  v_x= & (x+l/2)v/H_\rho & \mbox{for } x<0 \nonumber \\
  & (l/2-x)v/H_\rho & \mbox{for } x>0
\end{eqnarray}
Again, we assume $v\tau\ll H_\rho$.

First we regard the case when the field is perpendicular to the plane of the 
motions, see Fig.~\ref{fig:polpump}. 
This corresponds to the case of ``poloidal fields'' if our 2D flow 
pattern is identified with the extreme case of rotation-induced anisotropy
(meridional rolls or ``banana cells''). The field then behaves like a passive 
scalar; for each mass element, freezing-in requires $B/\rho=$const. Neglecting 
the horizontal flow we have $B(x,z;\tau )/\rho(z)=B_0/\rho(z-v_z \tau )$. 
(It is easy 
to see that the horizontal flow will only give a correction of order $v\tau
/H_\rho$.) With this
\[ {\cal E}_x =-\lang v_z B_y\rang = -B_0 \tau v^2 d_z\rho/\rho  \]
Writing this into (\ref{eq:avind}) we see that the pumping is directed 
downwards. The interpretation is again simple: downmoving elements get 
compressed, and this amplifies the field in the downflows relative to the 
upflows. The role of the horizontal flow is to return the decreasing up-moving 
flux into the downflows.

\begin{figure}[htbp]
  \centerline{\psfig{figure=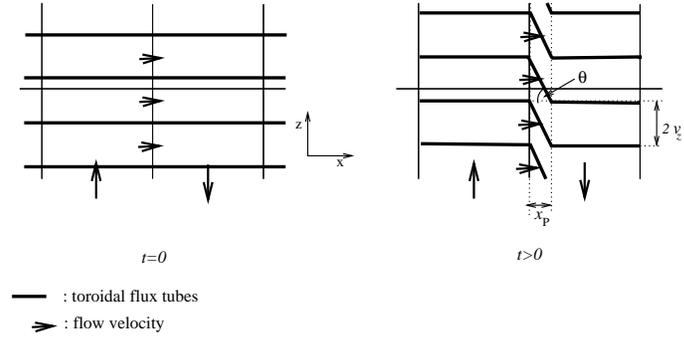,width=9cm,angle=90}}   	
  \caption[]{\label{fig:torpump} Density pumping for a toroidal field.}  
\end{figure} 

Next we consider the case of a ``toroidal'' field. The horizontal flow will 
in this case lead to the appearance of a vertical component of the fluctuating 
field in a finite part of the flow, 
as shown in Fig.~\ref{fig:torpump}. For the angle $\Theta$ one can work 
out
\[ \tan\Theta \simeq 4 H_\rho/l , \]
to first order in $v\tau/H_\rho$. With this
\[ {\cal E}_y =\lang v_z B_x -v_x B_z\rang =B_0/l\int^{l/2}_{-l/2} (v_z 
   +v_x\tan\Theta)\,dx \simeq -B_0 \tau v^2 d_z\rho/\rho  .  \]
The pumping is directed upwards. The interpretation is the following. The 
constant vertical velocity implies that in the upflows and in the central parts 
of the downflows the flux density does not change. In the peripheral parts of 
the downflows, however, where the field is tilted, the horizontal flow gives 
the 
tilted field an extra vertical lift, resulting in a reduced flux transport in 
the downflow as compared to the upflow.

The downwards directed pumping of a poloidal field by a 2D cellular flow was 
first pointed out by \cite{Droby:polpump}. \cite{Vainshtein:dens.pump} 
showed that for 3D isotropic turbulence the density pumping 
will vanish. Density pumping was recently studied by \cite{Kichat:dens.pump} 
under the assumption of quasi-homogeneity of turbulence.

\begin{figure}[htbp]
  \centerline{\psfig{figure=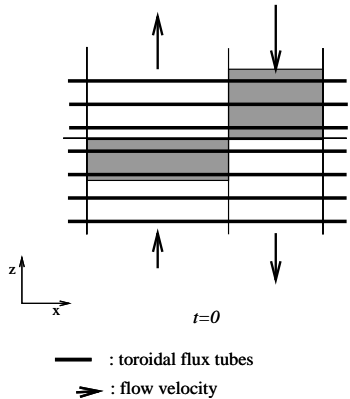,width=4.5cm,angle=90}}   	
  \caption[]{\label{fig:turpump} Turbulent pumping for a toroidal field.}  
\end{figure} 

\subsection{Turbulent Pumping}
The only inhomogeneous quantity is now $v^2$. As discussed in Sect.~2.4, in 
this case an asymmetry between up- and downflows is to be expected.
In the case of weak inhomogeneity, implied by (\ref{eq:assump}), we may use 
the linear relation 
$ 2 F_{\rm kin}\sim -\rho lv_t d{v_t^2}/dz $.
Comparing this with (\ref{eq:2compFkin}) we find $\Delta v=(v_u+v_d)/2\simeq 
2v\Delta f\sim -l dv/dz$ and $\Delta f=(f_u-f_d)/2=\Delta v/2v\ll 1$. Then all 
expressions can be expanded into a series to first order in $\Delta f$; in this 
approximation $F_{\rm kin}/\rho\simeq 2v^3\Delta f$. 

In the case of a poloidal field no turbulent pumping is expected to first order 
in $v\tau/H_{v^2}$, as the effect of the horizontal flow is again of higher 
order, and we are then back to the case of the density pumping of poloidal 
fields.

In the case of a toroidal field, with a manipulation similar to the previous 
cases (plus with an expansion in $\Delta f$), one can work out the result:
\[ {\cal E}_y =\lang v_z B_x -v_x B_z\rang = - B_0 \tau d_z{v^2}  . 
   \label{eq:turpump} \]
The pumping is directed downwards if the turbulent velocity decreases downwards.
The interpretation of (one contribution to) the pumping is shown in 
Fig.~\ref{fig:turpump}: as a consequence of the continuity equation and of a 
downward directed kinetical energy flux, the downflows are faster and they 
process a larger amount of flux through the reference level than do the upflows, 
resulting in a net transport downwards. (Another contribution is due to the 
action of horizontal flow on tilted flux tube portions, an effect analoguous to 
the density pumping of toroidal fields.)

Turbulent pumping was discovered by \cite{Zeldovich:turpump}, who christened 
the effect ``turbulent diamagnetism''. This expression is sometimes still used, 
although it is in fact a misnomer. Real diamagnetism is a dynamical effect, 
resulting from the orienting effect of the magnetic field on the medium, while 
turbulent pumping is a purely kinematic phenomenon. There is not even a 
mathematical similarity between the two effects: diamagnetism exerts its 
influence through the $\mu$ permeability, i.e.\ it is formally more similar to 
turbulent diffusion than to the pumping that involves an advective-type term.

Another confusion related to turbulent pumping is that its expression is often 
quoted in the form 
\[ \gamma =\nabla\beta\sim \nabla (v^2\tau) . \label{eq:badturpump} \] 
In the case when 
$\nabla\tau\neq 0$ this clearly differs from (\ref{eq:gammakif}) and from 
(\ref{eq:turpump}). For 2D turbulence the first equality in 
(\ref{eq:badturpump}) is indeed exact 
(\barecite{Cattaneo+:2Dturb.pump}) and it is also correct in 3D in order of 
magnitude. However, we should remember that the expression $\beta\sim\tau v^2$ 
is itself an order-of-magnitude formula only, so its differentiation will 
result in 
huge errors. The second (approximate) equality in (\ref{eq:badturpump}) 
should therefore be used with caution.


\section{Higher Order and Nonlinear Effects}
\subsection{Geometrical and Topological Pumping}
Geometrical pumping relies on the non-vanishing higher order correlation $\lang 
v_x\ptl_x v_z\rang$. The appearance of such a correlation implies that either 
the upflows or the downflows will have a mean horizontal divergence. The drag 
associated with these horizontal motions will then preferentially sweep the 
magnetic flux into the convergent component, which results in a net flux 
transport along the convergent component. For the sweeping effect to be 
significant in a correlation time $\tau$, a rather large (comparable to $v_z$) 
horizontal velocity component correlated with $\ptl_x v_z$ is needed, which 
requires a strong inhomogeneity by virtue of the continuity equation 
(\barecite{Petrovay:Becs}). Consequently, in turbulent flows this effect may 
only be important near a possible sharp boundary of the turbulent layer, and is 
therefore generally limited to low Reynolds number flows.

Topological pumping relies on a topological asymmetry of the up/downflows: in a 
horizontal cross-section of the layer, upflow areas may be connected, with 
isolated downflows or vice versa. A one-dimensional object like a field line 
will more easily be carried with the connected component. Of course, ultimately 
``what goes up must come down'', but, given sufficient time, magnetic 
diffusivity may be able to smooth out the resulting field into a mean field 
concentrated towards the direction of the connected flow. Again, the role of 
diffusivity implies that the effect will be more important for low magnetic
Reynolds numbers.

These effects were much studied and became undeservedly widely known in the 
seventies and early eighties 
(\barecite{Droby+Yuferev}, \barecite{Droby+:2nd}, \barecite{Arter:kinematic}, 
\citeyear{Arter:revisit}, \citeyear{Arter:dynamic}) when it was hoped that they may 
offer a solution to the buoyancy problem pointed out by 
\cite{Parker:buoy.prob}. The interest has dwindled after their unimportance was 
recognized in the high (magnetic) Reynolds number case. Recently, 
\cite{Petrovay:Helsinki} used topological pumping as an argument against a 
supposed sharp lower boundary of the solar convective zone.

\subsection{Negative Diffusivity and Small-Scale Dynamo}
\cite{Nicklaus+Stix} derived the expressions of $\mx{\alpha}$ and $\mx\beta$ 
for homogeneous turbulence in the fourth-order correlation approximation. One 
of their most interesting results is that the fourth-order correction to the 
diffusivity is negative, and it involves the helicity fluctuations. If the 
inhomogeneity of the field is strong, i.e.\ if $v_{\rm t}\tau\ll H_B$ does 
not hold, 
then this negative contribution may overcompensate the normal diffusivity.

A negative diffusivity implies that any small fluctuation will increase instead 
of being damped. In fact, no matter how large the overall scale of $\NB$ is, it 
may always develop infinitesimal small-scale fluctuations. The dominating 
negative term in $\beta$ for these small scales then means that these small 
fluctuations may reach large amplitudes before their growth is halted by 
nonlinear effects. This process does not influence the large-scale field, as 
the random fluctuations give no contribution to the net mean field; but it \it 
will 
\/\rm influence $\GB$. In effect, the consequence of the negative diffusivity 
at small scales will be the appearance of a \it source term\/ \rm in the 
transport equation for $\GB$, beside the transport terms known from the 
transport equation of $\NB$ (cf.\ Sect.~3.4 above). Turbulence closure 
computations (\barecite{DeYoung}, \barecite{Leorat+:smalldyn.closure}, 
\barecite{Durney+:basal}) as well as numerical simulations 
(\barecite{Meneguzzi+:smalldyn.simu}, \barecite{Meneguzzi+Pouquet:JFM}, 
\barecite{Nordlund+:undershoot}) indeed show this \it small-scale dynamo 
action. \rm On their basis, in the high (magnetic) Reynolds number case 
the source term may be phenomenologically approximated by
\[ \ptl_t\GB = \GB [1-\GB/(f_{\rm m} B_{\rm eq})] v_{\rm t}/l  , 
   \label{eq:smalldyn} \]
where $f_{\rm m}\sim 0.1$ is the magnetic filling factor at the nonlinear 
saturation.

Physically, the small-scale dynamo relies on the existence of small-scale 
helicity fluctuations in the flow which increase the unsigned flux density by 
stretch-twist-fold type motions.

\subsection{Magnetic Buoyancy}
Although buoyancy was much studied in the case of active fields, its 
description in MFT was for a long time limited to the use of phenomenological 
sink terms in the mean induction equation. Recently however a major 
breakthrough was achieved by \cite{Kichat+Pipin:buoyancy}. In their SOCA 
treatment of the problem they recover the expected result $\gamma\propto\NB^2$ 
for weak fields. For fields closer to equipartition however, the buoyant loss 
is significantly reduced by the Lorentz force.

\subsection{Density Self-Pumping and ``Turbulent Buoyancy''}
In Sect.~3 we have seen that anomalous pumping effects, e.g.\ density pumping, 
must rely on the anisotropy of the turbulence in a plane perpendicular to the 
pumping. The feedback of the magnetic field on the turbulence may itself induce 
such an anisotropy, thereby leading to the transport of the field. In 
particular, a horizontal anisotropy induced by a non-vertical magnetic field 
should lead to the vertical transport of the field. This problem was recently 
studied by \cite{Kichat+Rudiger:self.pump}. For strong fields they found a 
pumping downwards, in line with the expectation (the field should encourage 
turbulence in rolls around the field lines, which corresponds to the 
``poloidal'' case depicted in Fig.~\ref{fig:polpump}). Surprisingly, however, 
for weak fields the pumping was found to be directed upwards. The origin of 
this curious phenomenon of ``turbulent buoyancy'', as called by the authors, 
is unclear.

\subsection{Other Effects}
The above list is far from being exhaustive, even if we only regard the 
transport effects studied to date. As we go to higher orders, the number of new 
effects increases (although their importance will decrease, unless the 
inhomogeneity is very strong). Other effects are mentioned e.g.\ in 
\cite{Krause:compr.advection}, \cite{Moffatt:transport}, \cite{Nicklaus+Stix}.

An interesting but neglected  aspect of the 
transport problem is the role of \it vorticity\/ \rm in the magnetic field 
transport. The association of vorticity with narrow downflows and its 
interaction with flux tubes (\barecite{Nordlund+:undershoot}) suggests that a 
study of this problem would be worthwhile.


\section{Solar Applications}
\subsection{General Assessment and Convection Kinematics}
Numerical simulations (\barecite{Chan+Sofia:Science}) show that
in a quasi-adiabatically stratified turbulent convective layer the $l$  
correlation length is close to the $H_P$ pressure cale height. The scale 
heights of other mean quantities ($\rho$, $v_{\rm t}^2$, $\NB$ etc.) are 
usually greater than $H_P$, so in the deep solar convective zone the condition 
$l<H$ (though not $l\ll H$) is well satisfied, and the effects 
of first order in $l/H$, discussed in Sect.~2,  may be 
expected to dominate in the mean field equation (with the notable exception of 
the small-scale dynamo source term in the equation for $\GB$, discussed in 
Sect.~3.2 above). On the other hand, near the boundaries of the convective 
zone (e.g.\ in the photosphere) we may have $l\gg H$ and higher order effects 
may become important.

But what is the relative importance of different first order effects? In 
particular, is there any region in the convective zone where the importance of 
dynamo terms is negligible, so that our discussion of ``sterile'' transport in 
a parity-invariant flow applies? As the expressions of the terms appearing in 
$\vec{\cal E}$ involved velocity correlations of different orders, in order to 
answer this question a model of the solar convective zone (SCZ) with a reliable 
description of convection kinematics is needed. Unfortunately, the available 
models of solar convection were developed either for the needs of stellar 
evolutionists, who only require an average stratification (and perhaps the 
extent 
of the overshoot) or for the needs of solar/stellar granulation observers, for 
whom nothing short of a full 3D numerical simulation is good enough. Dynamo 
theorists, with their medium-level requirements are not well served by 
contemporary convection theory. Nevertheless, the last decade has brought a 
number of advances in the difficult problem of convection kinematics. 
This field would be worth a review in itself, so here we only recall some 
results more directly relevant to our present problem.

The best available model of the SCZ that can be recommended for use by dynamo 
theorists is the model by Unno, Kondo and Xiong (\barecite{UKX}, 
\barecite{Unno+Kondo}); in what follows: the UKX model. This is a fully 
non-local model; its most important assumption is that 
third-order correlations are proportional to the gradients of second-order 
correlations; this is tantamount to the assumption of weak inhomogeneity. 
More restrictive from the point of view of magnetic field transport theory is 
the assumption of a constant radial anisotropy ($\lang v_r^2\rang =2\lang 
v_\theta^2\rang =2\lang v_\phi^2\rang$). 
As we have seen in Sect.~2, the anisotropy of turbulence plays an important 
role in the transport. \cite{Petrovay:GAFD} computed the anisotropy of low 
Prandtl number turbulent convection in the homogeneous (Boussinesq) case. 
Inhomogeneity may 
however significantly modify the anisotropy. In a later work, 
\cite{Petrovay:Becs} evaluated the full anisotropy under the (invalid) 
assumption that inhomogeneity only influences the part of horizontal velocity 
that is correlated with the radial motions. In a complementary work, 
\cite{Kichat+Rudiger:aniso} computed the \it full \/\rm anisotropy caused by 
inhomogeneity in the case of quasi-isotropic original turbulence; this latter 
work also incorporates the effect of rotation. Although a complete solution of 
the anisotropy problem is still not around, the available results suggest that 
the anisotropy (both radial and horizontal) in the SCZ is significant, but not 
extreme.

\begin{figure}[htbp]
\vskip 6.5 cm
\hskip 2 cm
\includegraphics{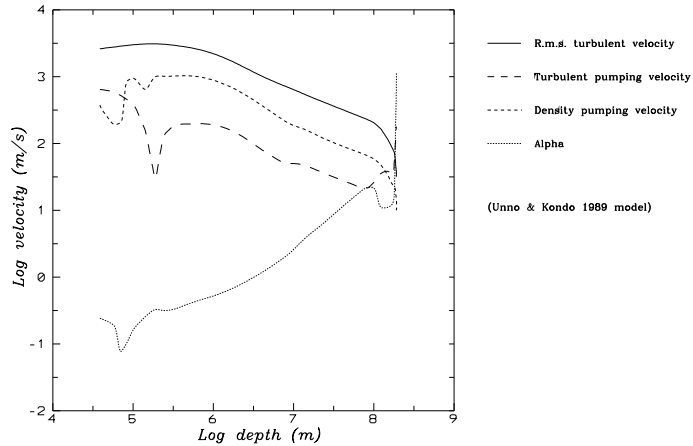}
\vskip -0.3cm  
  \caption[]{\label{fig:speeds} Characteristic velocites as functions of depth 
  in the solar convective zone. Note that only an the \it upper limit 
  \/\rm on $\alpha$ compatible with the given kinetic energy and an \it upper 
  limit \/\rm on the density pumping velocity corresponding to the unlikely 
  case of maximal anisotropy (meridional rolls) are displayed.}  
\end{figure} 

Fig.~\ref{fig:speeds} shows the characteristic velocities corresponding to 
different terms in the mean field transport equation as functions of depth in 
the SCZ. The relative magnitudes of the velocities characterize the relative 
importance of these terms. It is apparent that with the exception of the bottom 
of the SCZ the $\alpha$ term is by orders of magnitude smaller than the 
transport terms.  The characteristic 
velocity difference due to the shear by differential rotation (not shown) has a 
similar behaviour. This implies that with the exception of the bottom of 
the SCZ the dynamo terms are indeed negligible in comparison with the transport 
terms: \it in a simplistic treatment the solar dynamo may be split into a thin 
generation layer 
at bottom and a deep turbulent zone above; the distribution of the field 
generated in the dynamo layer will be determined by a dynamical equilibrium of 
turbulent transport processes in the convective zone. \rm Of course we should 
keep in mind that part of the magnetic field intruding into the convective zone 
from the dynamo layer below will be in active form (e.g.\ the thick flux loops 
giving rise to AR). The decay of such thick active tubes may act as a local 
source of passive fields in the convective zone. The amplitude of this 
\it active source \/\rm may be determined from observations of AR decay at the 
surface; but in deeper layers it presents a large uncertainty in our 
discussions of passive field transport. 

As mentioned above, higher order effects may come into play near the boundaries 
of the SCZ. This is likely to be the case in the photosphere where $l\gg H_P$, 
but unfortunately very little is known about flux transport in these 
circumstances. In an ongoing effort (\barecite{Faurob+me}) the observed 
photospheric height-dependence of $\GB$ is compared to vertical transport 
equilibrium models (as yet only including first order effects). In agreement 
with observations, these models indicate that $\GB\la 1$\,mT at the top of the 
photosphere. Whether such a low magnetic energy density is sufficient to 
explain the observed basal chromospheric Ca\,II\,K flux, as proposed by 
\cite{Durney+:basal}, remains to be seen.

The lower boundary of the SCZ presents an even more vexing problem. 
Phenomenological non-local mixing-length models traditionally predict a very 
sharp lower boundary with $v_{\rm t}^2$ suddenly dropping to zero in a thin 
layer. As pointed out by \cite{Petrovay:Helsinki}, in this case the extremely 
low velocity scale height should lead to strong geometrical and topological 
pumping. The flow topology in this layer is expected to be $d$\/-type in the 
notations of \cite{Petrovay:ApJ}, so all flux would be removed upwards from the 
dynamo layer. This contradiction seems to make it unlikely that the SCZ has a 
sharp lower boundary. In the models, the sharp bottom is a consequence of the 
\it a priori \/\rm assumption of a good correlation between velocity and 
temperature fluctuations which is not granted in the overshooting layer. 
Indeed, the more realistic UKX model shows a much more gradual decrease of 
$v_{\rm t}^2$ at the lower boundary with $H_{v^2_{\rm t}}\sim H_P$. This result 
agrees with numerical experiments (\barecite{Ludwig:Freibg}, Brandenburg, 
private comm.) and helioseismological constraints 
(\barecite{Monteiro+:seism.undershoot}). The problem is further complicated by 
the influence of magnetic fields and differential rotation on the structure of 
this important layer of the Sun.

\subsection{Vertical transport}
Fig.~\ref{fig:speeds} shows that the most important vertical transport 
effects are turbulent diffusion and turbulent pumping. The vertical 
distribution of passive magnetic flux in the SCZ should therefore be determined 
by a dynamical equilibrium of these processes, possibly influenced by density 
pumping and/or an active source. As $v_{\rm t}$ has a maximum at a shallow 
depth of 
200\,km in the UKX model, we may expect that the downwards directed turbulent 
pumping will ``press'' the field down to the bottom of the convective zone as 
far as turbulent diffusion, trying to ``smooth out'' the distribution, will 
allow. If for simplicity we regard a 1D plane parallel model with $\lang 
B_y\rang =0$, the only non-trivial component of the transport equation 
(\ref{eq:plantransp})--(\ref{eq:planpol}) is
\[ 0=d_z(\beta_{zz} d_z\lang B_x\rang ) +d_z[(\gamma_z +\tilde\gamma_z) 
   B_x] , \label{eq:NB} \]
with coefficients given by (\ref{eq:betakif})--(\ref{eq:tildegammakif}).

The vertical passive magnetic flux transport through the SCZ in a formulation 
similar to (\ref{eq:NB}) was apparently first considered by 
\cite{Droby:polpump}. By analogy with the density pumping of a poloidal 
magnetic field in 2D convection he proposed that the vertical transport is 
determined by a strong density pumping downwards. Soon thereafter, 
\cite{Vainshtein:dens.pump} showed that vertical density pumping will vanish 
for horizontally isotropic turbulence, and \cite{Schussler:vort.pump} stressed 
the importance of turbulent pumping in the vertical transport. 
The confusion concerning the correct expression of turbulent pumping, discussed 
above (end of Sect.~2.7), however led many to the false belief that turbulent 
pumping is negligible in the SCZ. As a consequence of this, 
density pumping made a comeback in the first quantitative study of the problem 
by \cite{Spruit+:Uloop}. The heuristic transport equation used in this work was 
in effect equivalent to (\ref{eq:NB}) containing only a (downwards) 
density pumping term and turbulent diffusion. An approximate analytic solution 
was also presented. Kichatinov's (\citeyear{Kichat:dens.pump}) study of the 
density pumping was applied to the SCZ by \cite{Krivod+Kichat}. 

\begin{figure}[htbp]
\vskip 6.5 cm
\hskip -0.8 cm\includegraphics{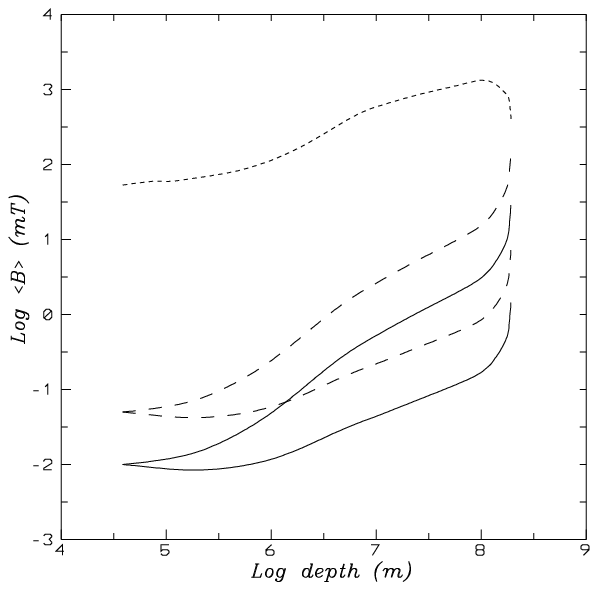}
\hskip 6.4 cm \includegraphics{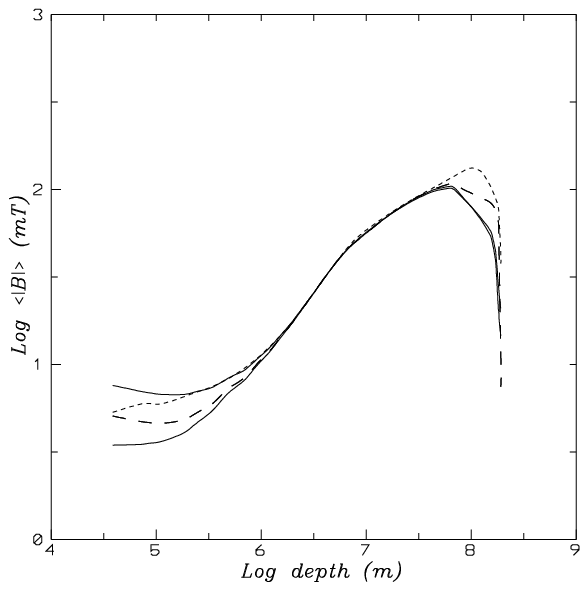}
\vskip -0.2cm  
  \caption[]{\label{fig:NBGB} Mean net flux density (left) and mean 
  unsigned flux density (right) as functions of depth in the SCZ. 
  Short dashes denote $B_{\rm eq}$ (left) and $f_{\rm m}B_{\rm eq}$ (right);
  other curves are solutions with different boundary conditions (see 
  \barecite{Petrovay+Szakaly:AA1}, for a detailed legend).}
\end{figure}

The somewhat speculative early models came into a different light with the new 
observational results summarized in Sect.~1.2 above. In order to see what these 
empirical constraints imply for the vertical distribution of passive magnetic 
flux, \cite{Petrovay+Szakaly:AA1} solved equation (\ref{eq:NB}) using boundary 
conditions taken from the observations and including turbulent pumping and 
turbulent diffusion, but neglecting the contribution due to the third and 
fourth terms in 
(\ref{eq:4terms}). \cite{Petrovay:Freibg} extended this work by including the 
neglected terms and the effects of sphericity (in an approximation that still 
preserved the mathemtical one-dimensionality of the problem). Equation 
(\ref{eq:NB}) can be integrated to yield
\[ \beta_{zz}d_z\lang B_x\rang +(\gamma_z+\tilde\gamma_z)\lang B_x\rang =-L_{\rm m}  , \]
where the $L_{\rm m}$ integration constant is the net flux loss from the Sun 
and it must obviously lie between 0 and $v_{\rm t,s}\lang B_x\rang_{\rm s}$ 
(`s' index refers to values near the surface, `b' to those at the bottom of 
the SCZ). Observations show that outside unipolar network areas $\NBx z$ is 
typically order of 0.01--0.1\,mT; this may also be applied to $B_x$ at some 
depth below the surface, assuming that averaging over the solar surface the 
ratio $|\NBx x |/|\lang B_z\rang |\sim1$. 
Figure~\ref{fig:NBGB} (left) shows the numerical 
solutions of the transport equation with different boundary conditions.
It is apparent that the flux density increases by several orders of magnitude 
to the bottom of the SCZ, confirming our expectation. Turbulent pumping 
therefore has the important role of ``helping'' the dynamo by sweeping the 
escaping flux back to the dynamo layer very effectively. It is also apparent 
that a relatively low flux density of $\sim 100$\,mT at the bottom will lead to 
a surface field comparable to that observed. Indeed, it is quite possible that 
all the net flux observed on the surface originates from passive fields in the 
dynamo layer by direct diffusion. On the other hand, we must remember that 
active source terms (neglected in (\ref{eq:NB})) may also be a significant 
source of net field. The observed rate of flux emergence in AR is too low to 
significantly modify the distribution (dynamical equilibrium is reached on a 
timescale $\sim 20$ days, comparable to the turbulent correlation time in the 
deep SCZ!) but AR may still be an important source of flux on a long term. In 
this latter case, the flux input by AR would be reprocessed by turbulent 
transport through the whole convective zone and would reappear again and again 
in a passive form. Finally, the possibility cannot be excluded that a much 
stronger active source 
operates in the deep SCZ in the form of a large number of ``failed active 
regions'' (\barecite{Petrovay+Szakaly:AA1}): rising flux loops decaying 
before they could make it to the surface.

An equation similar to (\ref{eq:NB}), with the added source term 
(\ref{eq:smalldyn}) should determine the equilibrium distribution of the 
unsigned flux density $\GBx x$. The integration of this equation 
(\barecite{Petrovay+Szakaly:AA1}, \barecite{Petrovay:Freibg}) yields the 
results shown in Fig.~\ref{fig:NBGB} (right). Note the weak dependence on the boundary 
conditions applied and note that the surface value of the unsigned flux density 
(2.5--7.5\,mT)
is \it not \/\rm a boundary condition here but an output in nice agreement with 
the observations of \cite{Faurob}. In fact the results show that even the 
resolved unsigned passive flux density of 0.5--1\,mT (cf.\ Sect.~1.2) is 
incompatible with a transport equation without source term; this can be 
regarded as indirect evidence for the operation of a small-scale dynamo mechanism in the 
Sun.

\subsection{Horizontal and 2D Transport}
The idea that horizontal transport of passive fields by meridional 
circulation and/or turbulent diffusion may have an important role in the 
solar dynamo 
goes back to \cite{Babcock:merid.circ} and \cite{Leighton:diffusion} who 
proposed that the poleward drift of $f$\/-polarity background fields at high 
latitudes is due to these effects. These two effects were combined into a 
single model by \cite{Sheeley+:first.combined} and further developed until a 
detailed comparison with synoptic magnetic maps yielded very positive results 
(\barecite{Wang+:1D}). Despite their success in interpreting the synoptic maps, 
these 1D horizontal transport models seem to be in conflict with the results of 
\cite{Stenflo:Bab-Leigh}, discussed in Sect.~1.2, which imply that the 
large-scale field pattern should be refreshed on a timescale of 3--30 days. 
Indeed, as we have seen, the vertical transport models yield a vertical 
transport timescale of $\sim 20$ days: over this time, the flux is vertically 
processed through the whole of the SCZ. The contradiction may disappear if we 
assume that the horizontal transport processes regarded by \cite{Wang+:1D} act 
throughout the whole SCZ. This would give a natural explanation for the 
remarkably low diffusivity (600 km${}^2$/s, much lower than $\tau v_{\rm 
t}^2\sim 10^4$\,km$^2$/s for the supergranulation) required by the comparison 
with the synoptic maps: this is about the value of $\tau v_{\rm t}^2$ in 
the deep SCZ.


In order to check whether the results of the horizontal diffusion models remain 
valid, detailed 2D (horizontal+vertical) models of magnetic flux transport 
are clearly needed. A first step in this direction was made by 
\cite{Wang+:1.5D} where the vertical dimension was included by a kind of 
vertical truncation of the equations. In Budapest we are presently working 
on a fully 2D model of the field transport; this model involves the 
numerical solution of equations (\ref{eq:spheri}--\ref{eq:spheritor}) 
with proper boundary 
conditions. Preliminary results from a fast implicit 
integration of a simplified, approximate form of the poloidal equation on a 
uniform $200\times 1000$ grid show that the 
behaviour of the solution qualitatively agrees with the expectation, i.e.\ 
turbulent  pumping presses down the field. 

A potential use of 2D transport models is to forge a link between the field 
distribution seen on the surface (as represented by synoptic maps or spherical 
harmonic expansions) and the field in the dynamo layer, thereby imposing an 
observational constraint on dynamo models. But the transport may also turn out 
to have a more fundamental role than this. If the 1D results of \cite{Wang+:1D} 
are confirmed in 2D, then the ``poleward branch'' of the butterfly diagram is 
fully due to field transport effects, instead of being a dynamo wave. In fact, 
\cite{Kichat:dens.pump} has proposed that transport effects may interpret \it 
both \/\rm branches of the butterfly diagram: the horizontal density pumping 
should transport the toroidal field equatorwards, the poloidal field polewards. 
Clearly, the full role of flux transport mechanisms in the dynamo can only be 
judged after full 2D transport models become available.

\vspace{3ex} \footnotesize \noindent 
{\bf Acknowledgements.}
I thank Rob Rutten for inviting me to the workshop. I am grateful to NATO 
and to the Soros Foundation for their financial support.
 
 \references
 
\end{document}